# Colorful Optical Vortices with White Light Illumination

Hongtao Wang[1, 2], Hao Wang[1], Qifeng Ruan[1], John You En Chan[1], Wang Zhang[1], Hailong Liu[3], Soroosh Daqiqeh Rezaei[1], Jonathan Trisno[4], Cheng-Wei Qiu[2, ]*, Min Gu[5, 6], Joel K. W. Yang[1,3, ]*

[1]Engineering Product Development, Singapore University of Technology and Design, Singapore 487372, Singapore

[2]Department of Electrical and Computer Engineering, National University of Singapore, Singapore 117583, Singapore

[3]Institute of Materials Research and Engineering, A*STAR (Agency for Science, Technology and Research), Singapore 138634, Singapore

[4]Institute of High Performance Computing, A*STAR (Agency for Science, Technology and Research), Singapore 138632, Singapore

[5]Institute of Photonic Chips, University of Shanghai for Science and Technology, Shanghai 200093, China

[6]Centre for Artificial-Intelligence Nanophotonics, School of Optical-Electrical and Computer Engineering, University of Shanghai for Science and Technology, Shanghai 200093, China

The orbital angular momentum (OAM) of light holds great promise for applications in optical communication, super-resolution imaging, and high-dimensional quantum computing. However, the spatio-temporal coherence of the light source has been essential for generating OAM beams, as incoherent ambient light would result in polychromatic and obscured OAM beams in the visible spectrum. Here, we extend the applications of OAM to ambient lighting conditions. By miniaturizing spiral phase plates and integrating them with structural color filters, we achieve spatio-temporal coherence using only an incoherent white light source. These optical elements act as building blocks that encode both color and OAM information in the form of colorful optical vortices. Thus, pairs of transparent substrates that contain matching positions of these vortices constitute a reciprocal optical lock and key system. Due to the multiple helical eigenstates of OAM, the pairwise coupling can be further extended to form a one-to-many matching and validation scheme. Generating and decoding colorful optical vortices with broadband white light could find potential applications in anti-counterfeiting, optical metrology, high-capacity optical encryption, and on-chip 3D photonic devices.

Analogous to vortices found in nature, *e.g.,* tornado, whirlpool, spiral nebula, and organism growth, the optical vortex discovered in the 1990s(*1*) attracts increasing interest in both classical optics and quantum science. The underlying property of light is the orbital angular momentum (OAM) that was first witnessed in Laguerre-Gaussian beams(*2*). Both state-of-the-art research and potential industrial applications benefit from this new degree of freedom to manipulate light, *e.g.*, optical communication(*3, 4*), optical manipulation(*5*), super-resolution imaging(*6*), high dimensional quantum computing(*7*), topologically complex electronic matter(*8*), and optical data storage and holography(*9-11*). To meet the increasing demand of diverse photonic devices, vortex beam generators and sorters were developed based on a variety of mechanisms(*12*), such as spiral phase plates (SPP)(*13*), metasurfaces(*14*), whispering gallery modes(*15-17*), spin-orbital interaction(*18*), holographic fork(*4, 19, 20*), and bound states in the continuum related physics(*21-24*). To increase information capacity and transmission speed, multiplexing of OAM channels has been widely exploited and shows great potential in optical communication in both free space and multimode fibers(*3, 10, 20, 25-31*).

OAM has yet to be explored for applications in information security and anti-counterfeiting, despite its promise in increased information capacity. In part, the requirement of spatio-temporally coherent light sources in generating vortex beams has hindered the incorporation of OAM in optical anti-counterfeiting(*32*). With the development of virtual/augmented reality and 3D displays, there is an increasing demand for full color display with white light illumination, instead of coherent sources, because they are readily available and potentially allow for compact and simpler implementation. Ambient white light is eye-safe, renders color to illuminated objects, and is closer to natural observation condition. Apart from visual display technologies, such as organic light emitting diodes (OLED) and white light holograms(*33, 34*), optical metrology also benefits from white light illumination and the suppression of phase and optical shot noise(*35, 36*). Moreover, compatibility with white light illumination is important for high-volume optical authentication and optical security.

To extend the generation of optical vortices to the broader ambient lighting conditions, we increase the spatio-temporal coherence of light sources through 3D printed free-form optical components. Partly due to the limitation of dispersion engineering and time-bandwidth product in 2D photonic system(*37*), previous planar metasurfaces and 2D photonic crystals have only exploited OAM at single or multiple discrete wavelengths(*10, 23, 38, 39*). Optical vortices that operate with incoherent and broadband ambient/white light sources are yet to be demonstrated. On the other hand, 3D printed nanophotonics hold advantages over planar photonic systems by providing an additional degree of freedom in thickness variation(*37*). With 3D printed free-form optical components, we achieved the coloration of optical vortices that produce brilliant visual effects of colorful rings whose diameters depend on its OAM value or topological charges (*i.e.*, indices for helical wavefronts).

Encoding both color and OAM information in the form of colorful optical vortices, these 3D free-form optical components act as building blocks for an optical security label on a transparent substrate. The invisibility of OAM information enhances the security towards next generation optical anticounterfeiting. The printing of these colorful rings can be achieved by spatially arranging these colorful optical vortices, and the topological charges can be examined at the designate focal plane under the illumination of broadband white light source. Alternatively, a pair of such optical elements with complementary topological charges and matching colors will provide an avenue for encoding and decoding discretized information of topological charges and colors. Thus, color, pattern placement, and OAM information can be encoded onto a 3D printed photonic "tally" for pairwise authentication.

Instead of employing a single photonic chip with optical cryptograph, pairwise photonic devices with complementary features hold the potential for more secure optical anticounterfeiting and ease of end-user validation. However, the development of such pairwise optical anticounterfeiting labels that do not necessitate public keys or additional arithmetic is still at its infancy(*32, 40, 41*). Here, we present a new pairwise optical anticounterfeiting device composed of colorful optical vortices with information encoded in their spectrum, OAM, and spatial coordinates. We draw inspiration from the multi-dimensional complementary features first introduced in the tiger tally of ancient China, one of the earliest authentication methods in the history of security and anticounterfeiting (Supplementary Information).

## Results and Discussion

The spatio-temporal coherence of light sources is represented by the complex degree of coherence $\gamma_{12}(\tau)$. It can be subdivided into the spatial $\mu_{12}$ and the temporal $\gamma(\tau)$ degrees of coherence, as shown in Fig. 1(a) (see details in Supplementary Information). To date, the generation of optical vortex beam necessitates coherent laser sources, shown near the origin of the coherence diagram (Fig. 1(a)). The resulting polychromatic and featureless focal spot prevent optical vortex beams to form from the broader region of incoherent light sources, especially for the most common and convenient ambient light source, *i.e.*, white light illumination. To expand vortex beams to incoherent white light illumination, improvements in both spatial and temporal coherence are necessary to achieve optical vortices instead of a blurry focal spot (Fig. 1(b)). The spiral phase plate (SPP) is miniaturized to 25 μm in diameter to achieve sufficient spatial coherence (Supplementary Information). On the other hand, the temporal coherence or the coherence length is enhanced by 6 to 13 times at by incorporating color filters in the form of nanopillars on the top surface of the SPP (Supplementary Information). As a result, colors are imparted to optical vortex beams. As shown in Fig. 1(b), our colorful vortex beam (CVB) unit operates under white light illumination and converts the blurry white focal spot into a green doughnut-shaped intensity profile, characteristic of optical vortex beams. CVBs thus convey information in the dimensions of OAM and colors. Apart from the color information and OAM topological charge, the spatial position of CVBs provide an additional degree of freedom for encoding information. Thus, a photonic tally encrypting three dimensions of information, *i.e.*, color, OAM, and spatial position is realized.

Figure 1(c) shows the concept of the photonic tally. Photonic Tally A is the "master" or lock piece, here shown with 9 CVB units with predefined topological charges, color, and positions. When illuminated with a broadband white light source, it generates CVBs seen as colorful doughnut shaped intensity profiles at its focal plane, where the size of the doughnut represents the magnitude of the topological charge. On the other hand, Photonic Tally B1 consists of CVBs with matching colors and positions with respect to Photonic Tally A but with conjugated topological charges. When overlaid and aligned with this photonic tally "key", the vortex beams generated by the first piece is exactly "untwisted" by the second and projects bright colorful spots (not doughnuts) at the combined focal plane of the tally pair. In addition to this one-to-one pairwise coupling between the photonic tally pair, a one-to-many matching and validation scheme of photonic tally security labels can be further exploited. For instance, when the same master tally is paired with another key, Photonic Tally B2, the image at the focal plane switches to a set of colorful rings with increasing radii, *i.e.*, topological charge, towards the outer perimeter. Leveraging on both color and topological charge, Photonic Tally B3 is an example of a tally with some unmatched color channels in addition to topological charge, leading to colorful rings only at specific positions, while blocking the rest. The final set of patterns from B1, B2, B3 can thus be validated against a database.

To achieve a compact colorful doughnut shaped intensity at the focal plane and increase the signal to noise ratio (SNR) of the photonic tally pair (Fig. S3), we introduce a converging phase to our CVB unit. Adding a lens profile to the SPP with pillars brings the intensity profile from the farfield to the Fresnel region, allowing good alignment between the photonic tally pair and avoiding color evolution along the optical axis(*39*). As shown in Fig. 2(a), our high SNR CVB unit is composed of a converging phase, spiral phase, and nanopillars filter. The CVB unit thus integrates three different functions of focusing, vortex beam generating, and color filtering into a single optical element. The height profile of the CVB unit is given by $H(r) = \left( f - \sqrt{f^2 + r^2} - l_i \varphi \right) / \Delta n(\lambda_j) + h_0$, for a given focal length $f$ and topological charge $l_i$, designed for at the transmittance peak wavelength $\lambda_j$. Here, $\Delta n$ is the refractive index difference between photoresist IP-Dip and the surrounding air environment, $r$ and $\varphi$ are radial and azimuthal coordinates respectively, and $h_0$ is the minimum thickness of the CVB unit. Next, nanopillars of constant height are arranged on the top surface of the CVB unit, such as to avoid placing pillars at the central singularity or lateral discontinuity to achieve a uniform arrangement of nanopillars (Fig. S4).

To speed up the two-photon polymerization (TPP) 3D printing of the photonic tally and reduce the memory consumption, we adopted a strategy of two-density three-step nanoscale 3D printing for the inner region and surface of the CVB unit (see details in Methods and Fig. S5). Instead of directly converting 3D models into Nanoscribe's file format, we coded a script to print our CVB units in three steps. First, the inner region of the CVB unit was printed on a glass substrate at high speed and relatively rough spacing and hatching distance (Methods), providing adhesion to the substrate and shell structure of the CVB unit, which we called the low-density stage. Secondly, in the high-density stage, the top surface of the CVB unit was printed with high scanning density and slower speed, providing more accurate control of the CVB unit height profile as well as the helical wavefront (Methods). In the third step, nanopillars with constant heights were added onto the CVB unit surface in the order of height coordinates

(See Supplementary Information Video 1). After the two-density three-step laser printing, the sample is developed to remove uncured resin, followed by UV-curing to strengthen the polymer nanostructures and avoid obvious shrinkage caused by incomplete curing. As shown in Fig. 2(b) to (d), the tilted scanning electron microscope (SEM) images show the near perfect fabrication result of the CVB unit.

As the top surface of the CVB unit is curved instead of flat, we investigated the effect of slanted surface on the nanopillar color filters. Surface gradient of the CVB unit was calculated and converted to local surface angles (*i.e.*, the angle between the norm vector of the tangential plane and the vertical direction at corresponding surface point, depicted in Fig. 2(c) inset and Fig. S6). Compared to the diameter of nanopillars (400 nm), the surface gradient of the CVB unit is slowly varying (Fig. 2(d)), thus can be approximated as a constant slope locally beneath a single nanopillar. Spectra of nanopillars with different heights on tilted slopes were simulated using finite-difference time-domine (FDTD) and converted to color palette under the condition of CIE standard illuminant D65 and standard observer, as shown in Fig. 2(e). The color of nanopillars at a particular height gradually redshifts with increasing slope angles. The saturation and brightness of nanopillar colors also slightly decreased. The red color shift was caused by the extension of spatial period of the unmodulated planewave passing through the gaps between nanopillars. The corresponding vertical wave vector component along the height direction of nanopillars is modified to $k_{air} \cdot \cos\left(\arcsin\left(n_{IP-Dip} \sin(\theta)\right) - \theta\right)$, which is smaller than the case for perpendicular incidence. Consequently, the deflected energy flow and the decrease of intensity contrast between harmonic wave modulated by nanopillars and unmodulated harmonic wave led to a weakened interference, resulting in broader spectra and lower saturation of colors. Based on the simulation results, we can identify suitable nanopillar heights for the CVB units with different topological charges and colors. However, within a small slope angle range from 0° to 20°, the saturation and brightness of nanopillar colors remain relatively uniform if we choose nanopillar heights to be 1.0 μm, 1.4 μm, and 2.1 μm for blue, green, and red color channels respectively, as shown in Fig. 2(e) and (g). These nanopillar heights for the CVB units are consistent with those used recently in nanopillar-based color prints(*42*) and optical security devices(*43*).

We designed CVB units with 25 μm diameter and 60 μm focal length to account for multiple factors, *e.g.*, spatial coherence, numerical aperture (NA), SNR of the CVB unit, thickness, color uniformity, fabrication time, and ease of alignment of two tallies. The experimental averaged spectra (Fig. 2(f)) and optical micrograph (Fig. 2(g)) of the CVB unit verify these three distinct color channels, which can be recognized easily by eye or imaged with a charge-coupled device (CCD) camera followed by image postprocessing. As shown in Fig. 2(h), the topological and spectral dimensions can be decoupled using the CVB units, thus colorful optical vortices can be achieved by adjusting nanopillar heights and surface topography to generate a library of colorful rings with different topological charges. Combined with precise placement of these integrated optical elements, *i.e.,* CVB units, an optical motif with colorful dots with as-desired pattern can be realized on the sample surface plane based on TPP 3D printing (Methods). The magnitude of the OAM topological charge can be determined by measuring the color rings prints at the focal plane (Fig. 2(h)).

The coloration of optical vortices and their corresponding topological charges was further examined through experiments and numerical simulations. We built an optical setup (Fig. S7) to obtain the micrographs of the CVB unit (Fig. 2(g) and Fig. S8) and the doughnut shaped intensity of colored optical vortices (Fig. 2(h) and Fig. S9) under broadband white light illumination. With the modulus of topological charges increasing from 0 to 5, the radius of colorful rings in three color channels increases monotonically (Fig. S9 and Fig. S10). As a definitive test, the interference pattern between vortex beam and unaffected planewave can be obtained under laser illumination (Fig. 2(i) and Fig. S11). The difference value of interference fringes $n_{right} - n_{left}$ indicates the topological charge of incident vortex beam, which agrees well with phase wrapping along an arbitrarily closed loop $C$ in the focal plane, $q = \frac{1}{2\pi} \oint_C \nabla \varphi(\mathbf{r}) \cdot d\mathbf{r}$ (Fig. 2(i), Fig. S12 and Fig. S13). In short, both doughnut shaped intensity profiles and topological charges of our colored optical vortices are verified in both experimental and simulation results (More details of topological charges analysis can be seen in Methods and Supplementary Information).

We then investigated the spectral response and validate the effectiveness of the CVB unit in generating bright ring intensity profiles near its designed focal plane (60 μm away from the CVB unit surface). We performed FDTD simulations and extracted the beam intensity at the focal region for different wavelength channels (see details in Methods). In Fig. 2(j), the simulation results clearly show the color filtering of nanopillars with different heights and the topography of the CVB unit successfully converge these colorful optical vortices into the focal region, 55 μm to 65 μm away from the CVB unit surface. Within this focal region, we clearly observe the doughnut shaped intensities of three CVBs with distinct spectral bands corresponding to their respective transmittance spectra.

A single piece of fabricated photonic tally is shown in Fig. 3. The first column with red outline shows scanning electron microscope (SEM) images of three photonic tally samples. The CVB units are homologous, *i.e.*, sharing similar appearance, thus hiding both color and OAM information. When focused on the top surface of the CVB units, the optical micrographs of photonic tally samples in transmission mode exhibit as-designed colorful patterns, as shown in the second column with orange outline. Upon closer examination, one can notice that two of the samples are patterned with the letters "SUTD" in alternating red and green letters, while the third is patterned as "Tetris"-like blocks. To read out the OAM information, we adjust the microscope focus to the designed focal plane of the photonic tally, collimate the broadband white light source by reducing the aperture of illumination lamp, and increase the integration time of the CCD camera. The color prints of "SUTD" and "Tetris" then appear as brightly colored rings of different sizes (See Supplementary Information Video 2), showing both OAM and color information of these photonic tally piece. The topological charges in panel (a) encrypt the first 100 digits of "π" by subtracting each digit by 5 to map the numbers from [0, 9] to [-5, 4], *e.g.,* 3.1415 → -2 -4 -1 -4 0. In addition, a spatial pattern consisting of colorful rings arranged in an easily recognizable manner can also be achieved. The topological charges in panel (b) show an array of colorful rings whose magnitude of topological charges are constant around square contours "□" but increase outwardly from 1 to 5. These topological charges have positive values in odd quadrants and negative values in even quadrants as indicated. In addition to digital codes and spatial pattern, arbitrary spatial coordinates can be further exploited for encryption, as shown in Fig. 3(c).

To verify the concept of pairwise authentication of the photonic tally pair, we fabricated and tested the "key" tally for the "SUTD" motif. Here, the corresponding topological charges are conjugated. These charges account for the same degree of phase wrapping but in opposite handedness. Thus, the net topological charge will be 0 after passing through a pair of conjugated CVB units. If this works, combining the tally pairs will convert the previously colorful ring array into an array of colorful dots instead. This cancellation of topological charges and conversion from topological vortex beams to simple converging beams are shown in Fig. 4(a) and (b). We show this for the two "SUTD" colored photonic tallies with OAM encrypted digital codes for "π" and spatial pattern information of "□". The corresponding "keys" with CVBs shown in the middle column will untwist the "master" optical vortices and produce colorful dots at desired positions (Supplementary Information Video 3 to 4). Distinct from the conventional lock and key system, our photonic tally pair holds the advantages of reciprocity due to the reversibility of light. Switching the position of two complementary photonic tally pieces, the colors and positions of these focused colorful dots are unchanged. This reciprocity may find application in mutual optical validation.

To demonstrate the one-to-many matching and validation scheme, Fig. 4(a), (c) and (d) show that a single tally encrypting "π" manifests three different patterns of colorful rings at the combined focal plane when combined with different keys. With its conjugated piece "π*", the test results are purely colorful dots without topological nontriviality. However, when combined with photonic tally pieces "A" and "B", the test results show the recognizable spatial patterns of "□" and "✲" instead (Supplementary Information Video 5 and 6). The output topological charges in panel (d) show an array of colorful rings with constant topological charges in each sector that increasing anticlockwise. The topological charges of photonic tally pieces "A" and "B" was calculated according to the desired results. $S_2 = B\{T - S_1 \in C\} \cdot (T - S_1) + B\{T - S_1 \notin C\} \cdot (-T - S_1)$, where $S_1$ is the topological charge array of the first piece, $S_2$ is the calculated topological charge array of the second piece, and $T$ is the target topological charge pattern. $B$ is a binarization operator, while $C = [-L, L]$ is the integer set of available topological charges, where $L$ corresponds to the maximum available topological charge. This one-to-many matching and validation scheme broaden the content of pairing in photonic devices and introduces tremendous convenience to optical anticounterfeiting and cross optical validation.

To further utilize the color channel, we combine photonic tally pairs possessing the CVB units of unmatched colors at arbitrary positions (Fig. 5). The seemingly similar-looking CVB units show colorful rings of different sizes with different topological charges under the collimated broadband white light. When combined photonic tally pieces with conjugated topological charges (Fig. 5(a) and (b)), we can choose to completely block certain bright dots, *e.g.*, by combining red and blue filters. Only the cascaded CVB units sharing the same color can produce discernible colorful dots when combined with its pair. In contrast, the cascaded CVB units with different subtractive colors will significantly attenuate the detected signal on the CCD camera and result in a naught in corresponding positions (Fig. 5(d)). On this principle, the "Tetris" photonic tally piece will also eliminate some Tetris pieces (not according to the actual rules of the game) when combined with its companion Tetris of different colors. As shown in Fig. 5(c) and (d), only half of the Tetris pieces remain after cascading with an altered "Tetris" photonic tally piece (Supplementary Information Video 7). The elimination of colors and the annihilation of topological charges increase the diversity of our photonic tally pairs and improve the security level.

In conclusion, the TPP lithography enables the 3D printing of custom designed optical elements that achieve high spatio-temporal coherence to generate colorful vortex beams (CVB) under the white light illumination. When stacked and aligned in a pairwise manner, these CVB units generate and perform arithmetic addition functions of topological charges, in addition to transmission or blocking of the vortex beams with same or different colors. Encrypting spatial coordinates, color, and OAM information, our photonic tallies broaden the validation methods of optical security labels with a pairwise working principle. The resulting output patterns can be machine read and compared with a secured database for physical-digital authentication. By controlling colors and topological charge independently, we achieve 33 possible CVB states for each unit. With only a 10×10 array of such CVB units, each of our photonic tally holds up to $33^{100}$ possible combinations, ensuring unique tallies for next-generation optical anticounterfeiting devices.

# Methods

## Optical simulation of the CVB unit

The simulation of nanopillars on tilted substrate is carried out using a commercial software Lumerical FDTD. A 5 by 5 nanopillars array is simulated with perfect match layers (PML) boundary conditions. The calculated spectra are converted to far field and the energy within half collecting angle of 26.7° is collected, corresponding to a 20× magnification and NA = 0.45 objective lens used in the experiment.

The spectrum responses of the CVB unit in Fig. 2(j) are also simulated using Lumerical FDTD. the CVB unit with diameter of 25 μm and focal length of 60 μm is built with script and simulated with PML boundary condition. A series of power monitors are placed near the designed focal region and the intensity of focused vortex beam is normalized at each z position. Phase wrapping and intensity profiles in Fig. S12 and Fig. S13 are calculated from electric field on the focal plane.

## Two-density three-step nanoscale 3D printing

The fabrication of CVB units and photonic tally piece are conducted using Nanoscribe GmbH Photonic Professional GT based on two-photon polymerization 3D printing. A 63× objective lens with NA = 1.4 is used in dip-in laser lithography (DiLL) mode. Before laser printing, the glass substrate is treated using TI PRIME adhesion promoter (MicroChemicals GmbH, Germany). In the fabrication of inner region of the CVB unit, ContinuousMode of GalvoScanMode is used, ScanSpeed is 20,000 μm/s, LaserPower is 24% of the total power (50 mW) corresponding to 12 mW. The hatching and slicing distance are 0.2 μm and 0.3 μm, respecctively. In the fabrication of top surface of the CVB unit, ContinuousMode of GalvoScanMode is still used, ScanSpeed is 20,000 μm/s, LaserPower is 12 mW. The hatching distance is 0.1 μm, while the heights are defined by the profiles of CVB units. In the fabrication of nanopillars, PulseMode of GalvoScanMode is used, ExposureTime is 0.06 ms, LaserPower is 25 mW. The global PiezoSettlingTime is 10 ms and GalvoSettlingTime is 2 ms. After exposure, the sample is developed in propylene glycol monomethyl ether aceta for 5 mins, isopropyl alcohol for 5 mins, and methoxy nonafluorobutane for 7 mins. When developed in isopropyl alcohol, the sample is cured with UV light for 5 mins with 70% maximum power (DYMAX, MX-150, 405 nm).

**Optical characterization of the CVB unit and photonic tally**

The CVB units are tested using our homemade optical setup. When illuminated with halogen lamp and block the cyan light path, broadband white light (yellow light path in Fig. S7) passing through the sample and then be the 4f optical system (pink light path). The signal of focused CVB is detected by the CCD Camera and shown in Fig. S7. When illuminating using lasers with corresponding wavelengths and letting CVB (pink light path) interfere with planewave (cyan light path), the spatial interference patterns of the CVB units are captured and shown in Fig. S11.

Photonic tally pieces and photonic tally pair are characterized using a microscope (Nikon Eclipse LV100ND) in transmission mode. The broadband white light is collimated by reducing the aperture size of the halogen lamp. Tuning the focus of microscope away from top surface of photonic tally piece to the designed focal plane, indistinguishable color prints (second column with orange edges in Fig. 3) are converted to colorful rings with information of topological charges (third column with purple edges in Fig. 3). The alignment of photonic tally pairs and corresponding optical characterization are also performed on the same microscope.

## Associated Contents

**Online Content**

Any methods, additional references, source data, extended data, supplementary information, acknowledgements; details of author contributions and competing interests; and statements of data and code availability are available online.

**Data availability**

The data that support the figures and other findings of this study are available from the corresponding authors upon reasonable request. Source data are provided with this paper.

**Code availability**

The code used for the photonic tally design and characterization is available from the corresponding author upon reasonable request.

**Acknowledgement**

J.K.W.Y. acknowledges funding support from the National Research Foundation (NRF) Singapore, under its Competitive Research Programme award NRF-CRP20-2017-0004 and NRF Investigatorship Award NRF-NRFI06-2020-0005. C.-W.Q. acknowledges financial support from the National Research Foundation, Prime Minister's Office, Singapore under Competitive Research Program Award NRF-CRP22-2019-0006. C.-W.Q. is also supported by a grant (A-0005947-16-00) from Advanced Research and Technology Innovation Centre (ARTIC) in National University of Singapore. Min Gu acknowledges the support from the Science and Technology Commission of Shanghai Municipality (Grant No. 21DZ1100500) and the Shanghai Frontiers Science Center Program (2021-2025 No. 20).

**Corresponding Author**

*E-mail: joel_yang@sutd.edu.sg

*E-mail: eleqc@nus.edu.sg

**Author Contributions**

H.T. Wang, J. K. W. Yang and C. W. Qiu conceived the idea of colorful optical vortices and photonic tally pair. H.T. Wang performed the design, numerical simulation, fabrication, and characterization of photonic tally pair with assistance from H. Wang and drafted the manuscript. All the authors contributed to the data analysis and manuscript revision. J. K. W. Yang and C. W. Qiu supervised the whole project.

**Competing interests**

The authors declare no competing financial interests.

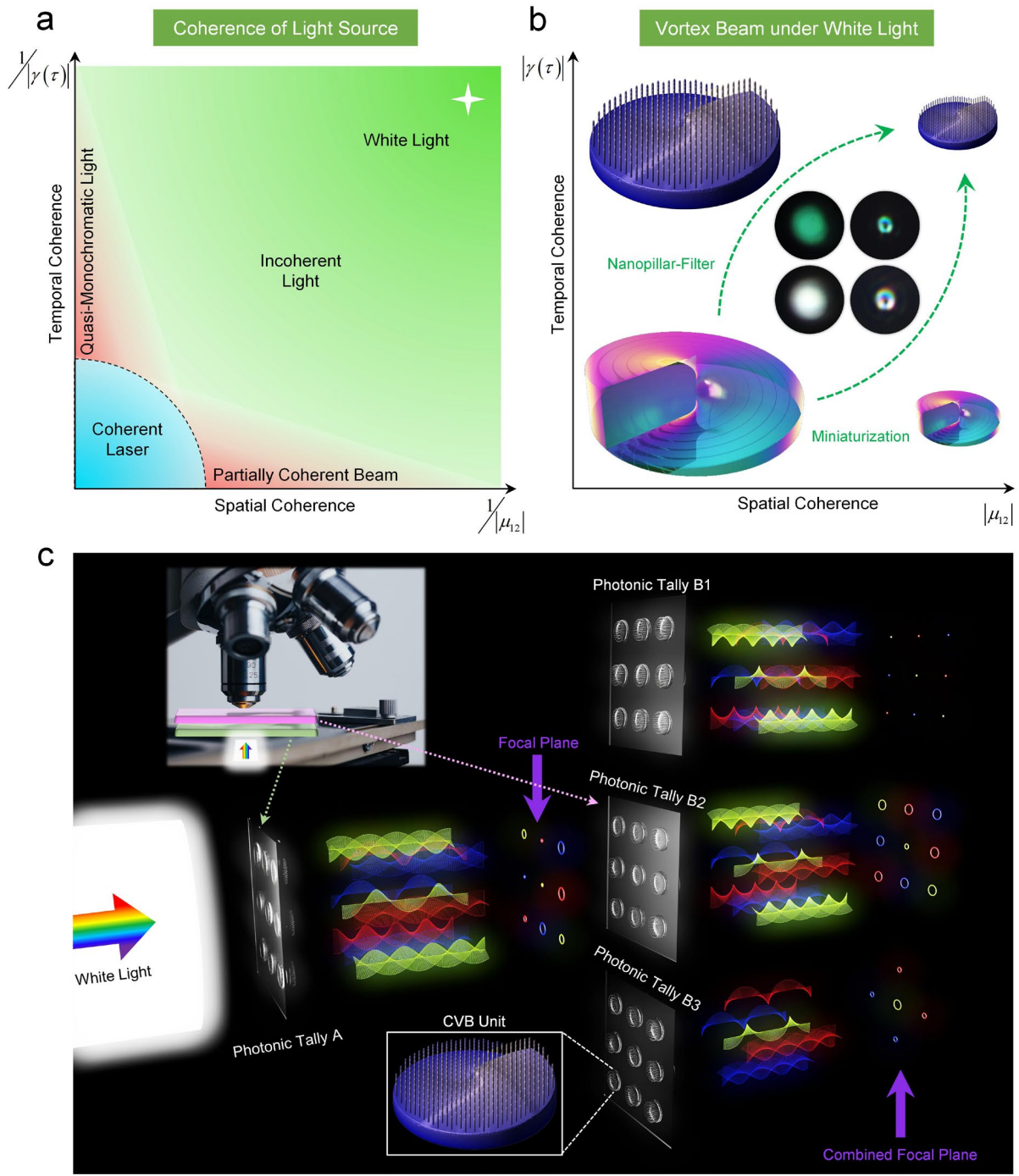

Figure 1. Schematic of colorful vortex beam (CVB) unit and photonic tally pair. (a) Spatio-temporal coherence diagram of light sources. The coherent laser near the origin exhibits the highest spatio-temporal coherence, while the white light source indicated with the cross is an example of poor coherence. (b) Four potential structures for generating CVB with white light illumination. The spatial coherence of incident white light is enhanced by miniaturizing the spiral phase plate (SPP), while the temporal coherence of incident white light is enhanced by adding nanopillar-based color filters. Through enhancements both in spatial and temporal coherence, the blurry white focal spot is converted into the green doughnut shaped intensity profile (insets). (c) Schematic of photonic tally pair. With white light illumination, the photonic tally forms an array of colorful rings at the focal plane, encoding color and OAM information. Combined with different photonic tallies B, distinct output patterns form at the combined focal plane, *e.g.*, colorful dots array, colorful rings with specific topological charges (*e.g.*, increasing outwardly), and eliminated color channels. Inset with white outline: schematic of colorful vortex beam (CVB) unit.

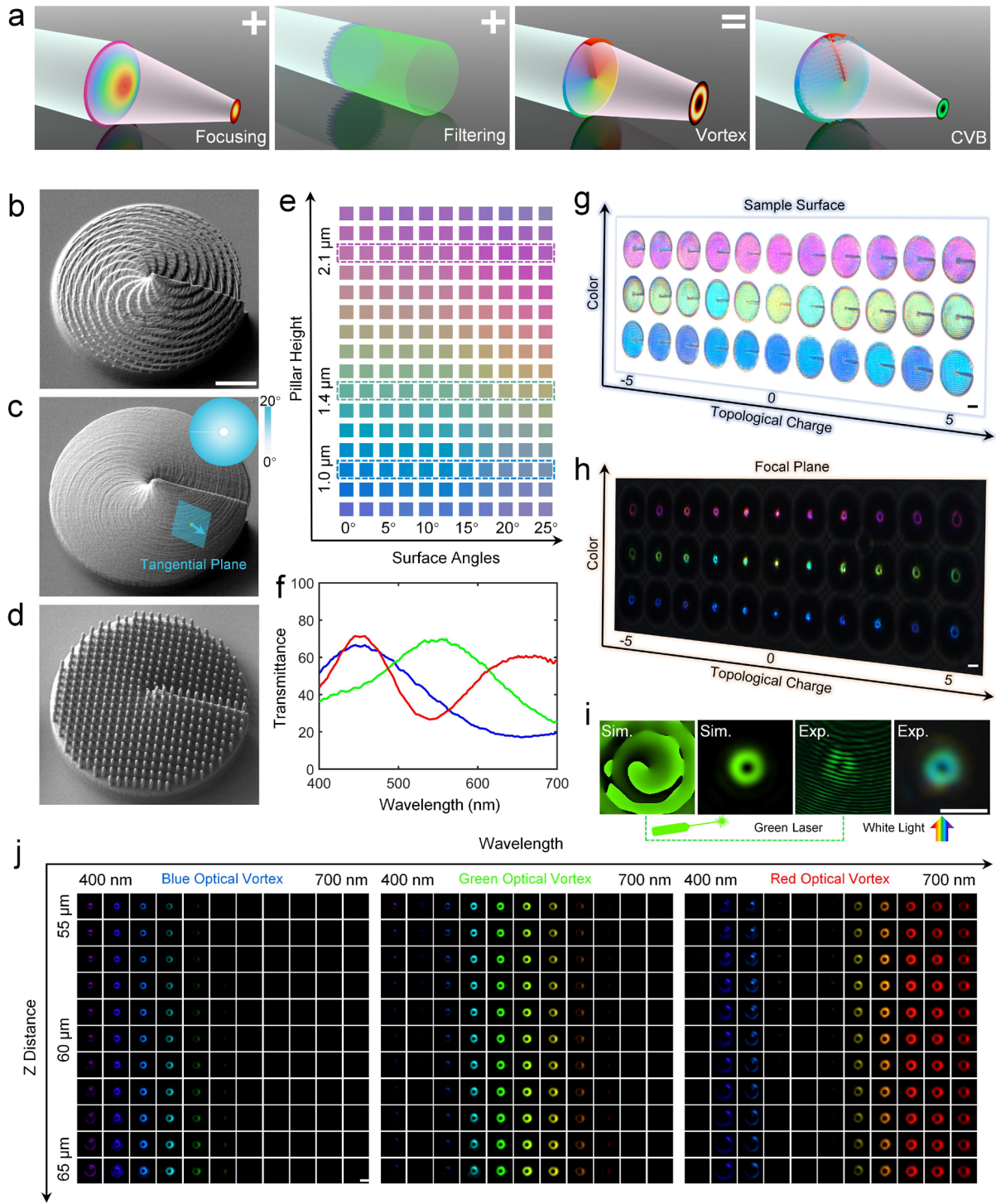

Figure 2. Design principle and experimental results of the CVB units. (a) Combination of converging microlens (I), nanopillar color-filters (II), and spiral phase plate (III) into a single optical element and the realization of a focused CVB. (b)-(d), SEM images of the CVB unit at different stages of our two-density three-step fabrication process showing rough inner region (b), smooth top surface of the CVB unit (c) and printing of nanopillars (d). (e) Simulated colors

of vertical nanopillars on surfaces with different slope angles. (f) Averaged spectra of the CVB unit in three color channels, marked in (e). (g) Optical micrographs of the CVB units in three color channels with topological charges ranging from -5 to 5 achieving 33 different possible states per CVB unit. (h) Optical micrograph of doughnut shape intensity profiles of the CVB units when the vortex beams are observed at focal plane under the white light illumination. (i) Simulated intensity and phase profiles of the converged optical vortex at desired focal plane; corresponding experimental intensity profile and interference pattern. (j) Spectral response of the CVB units, showing converging optical vortices under single wavelength illumination in the desired focal region. All scale bars are 5 μm.

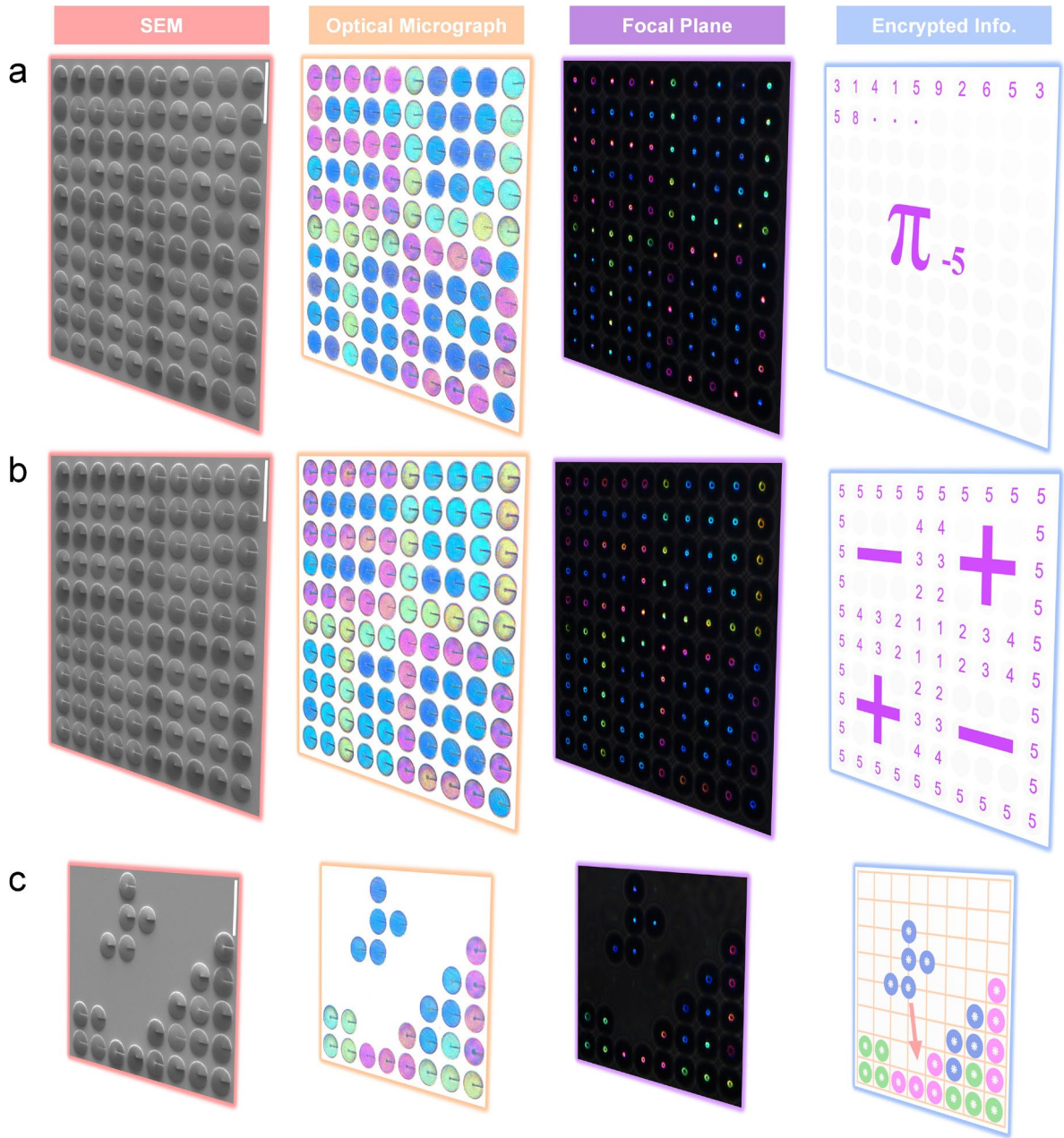

Figure 3. Single photonic tally piece and corresponding encrypted information. Photonic tally pieces showing "SUTD" letters in color and 100 first digits of "π" in topological charges (a), predesigned spatial pattern "□" (b). (c) Colorful tetris with arbitrary spatial coordinates and random topological charges. All scale bars are 50 μm. Column with red outlines is SEM images. Column with orange outlines shows optical micrographs of the fabricated photonic tally pieces. Column with purple outlines shows optical micrographs taken at the focal plane of photonic

tally pieces and illuminated with collimated broadband white light. The last column with blue outlines is the corresponding hidden topological charge information.

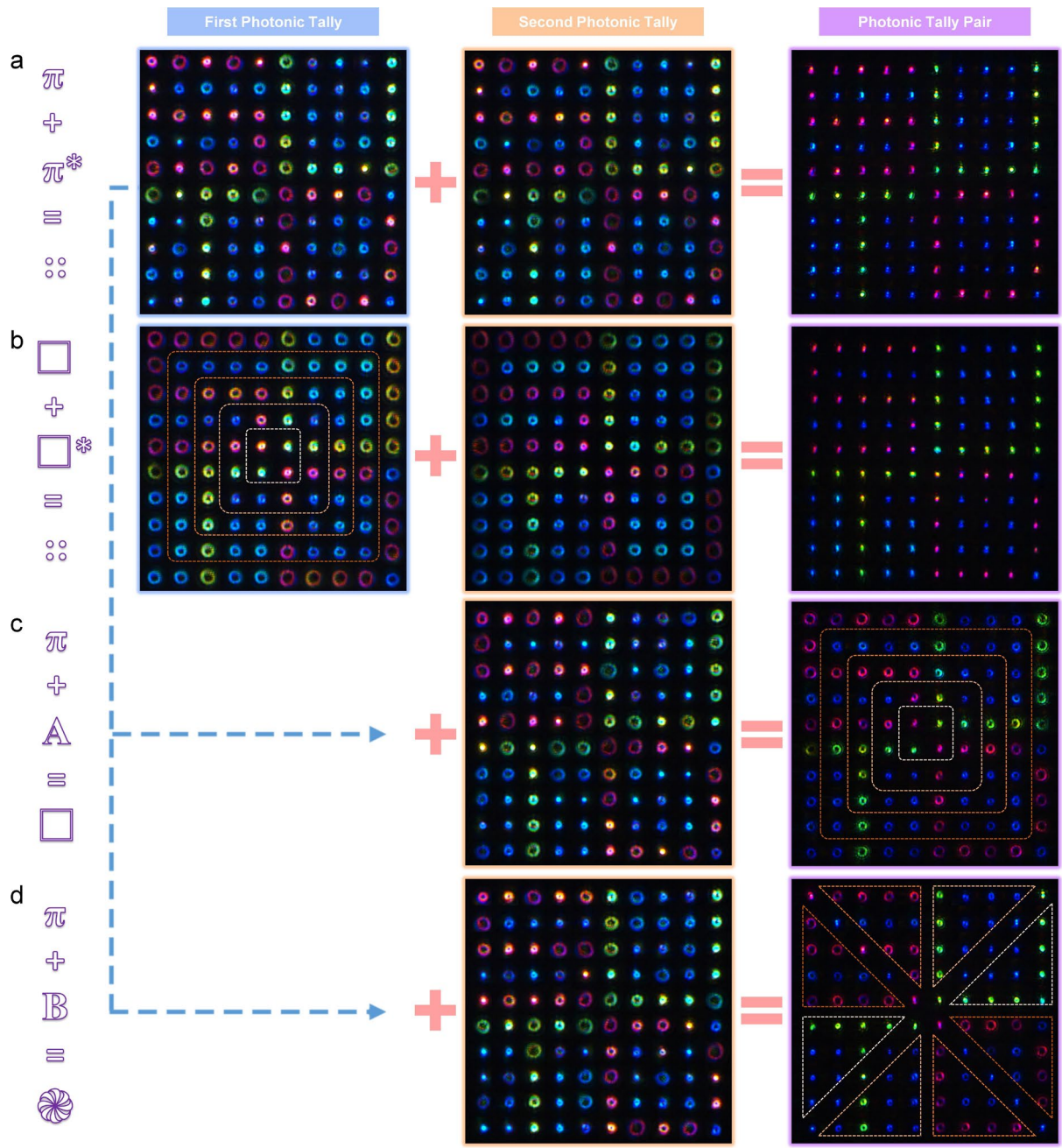

Figure 4. Combined effects of photonic tally pairs. (a)-(b) Conjugated photonic tally pairs encrypting information of "π", and "□". (a), (c) and (d), one-to-many matching and validation scheme of photonic tally pairs. Same photonic tally piece of "π", combined with three different photonic tally companion "π*", "A", and "B", manifest three distinct colorful ring patterns of colorful dots, "□", and "🌀". To exhibit the optical vortex array in this limited space, the micrographs were cropped to enlarge the size of the rings. Original micrographs without postprocessing are shown in Fig. S15.

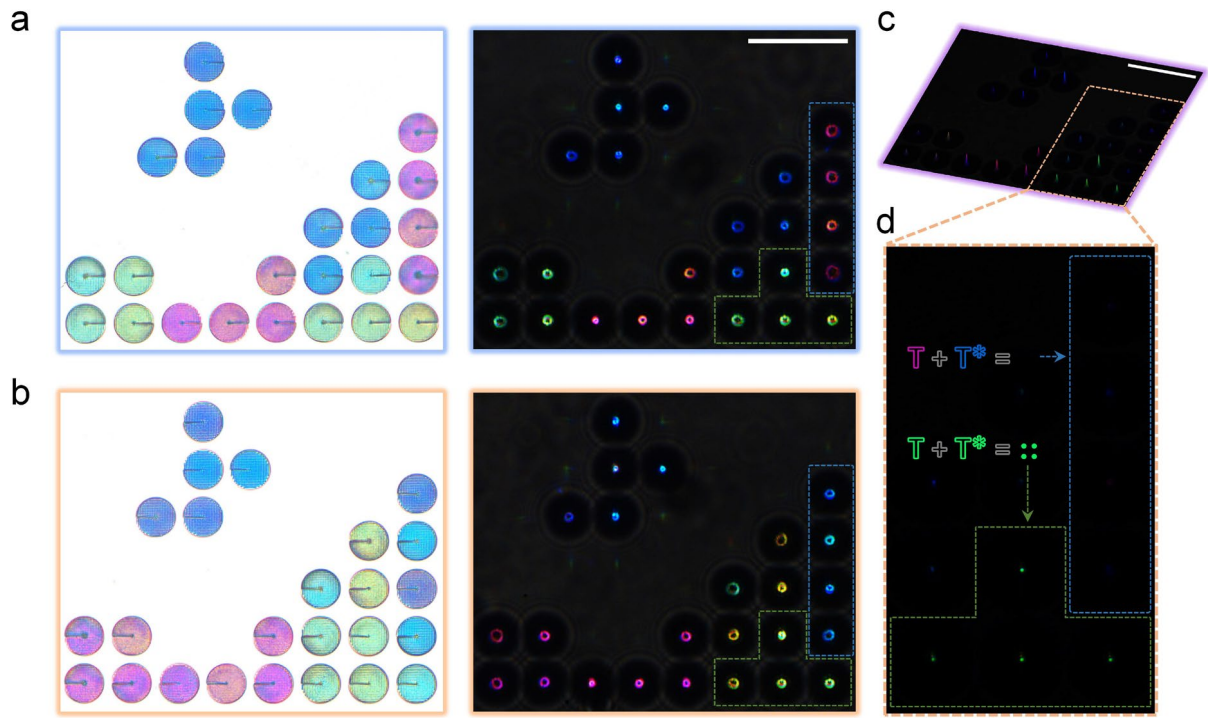

Figure 5. Photonic tally pair with different colors. (a) The first photonic tally piece, with three colors "Tetris" and random topological charges. (b) The second photonic tally piece. Left panels: optical micrographs of photonic tally pieces. Right panels: focused CVBs. (c) Combined optical effects of photonic tally pairs. (d) Enlarged image of dashed orange box in Fig. (c), "Tetris" with same colors turns out to be colorful dots (green dashed region), while "Tetris" with different subtractive colors indicates the elimination of different color channels (blue dashed region). All scale bars are 50 μm.

**Supplementary Information**

# Colorful Optical Vortices with White Light Illumination

Hongtao Wang[1,2], Hao Wang[1], Qifeng Ruan[1], John You En Chan[1], Wang Zhang[1], Hailong Liu[3], Soroosh Daqiqeh Rezaei[1], Jonathan Trisno[4], Cheng-Wei Qiu*[,2], Min Gu[5, 6], Joel K. W. Yang*[,1,3]

[1]Engineering Product Development, Singapore University of Technology and Design, Singapore 487372, Singapore

[2]Department of Electrical and Computer Engineering, National University of Singapore, Singapore 117583, Singapore

[3]Institute of Materials Research and Engineering, A*STAR (Agency for Science, Technology and Research), Singapore 138634, Singapore

[4]Institute of High Performance Computing, A*STAR (Agency for Science, Technology and Research), Singapore 138632, Singapore

[5]Institute of Photonic Chips, University of Shanghai for Science and Technology, Shanghai 200093, China

[6]Centre for Artificial-Intelligence Nanophotonics, School of Optical-Electrical and Computer Engineering, University of Shanghai for Science and Technology, Shanghai 200093, China

**The analogy between "tiger tally" and "photonic tally".**

The tiger tally is a 2,300-year-old tiger-shaped tally from ancient China, was used to authenticate military orders. One piece was held by the military general, and the other complementary piece held by the monarch. When brought together, the two pieces would precisely match like pieces of a puzzle, as their outer shapes, inner morphology, and surface characters are complementary, as shown in Fig. S1(a). Drawing parallels to the tiger tally, we designed two 3D printed photonic tallies to generate colorful optical vortices under broadband white light illumination. This "photonic tally pair" contains conjugated topological charges at corresponding spatial positions (Fig. S1(b)). Moreover, the colors of vortex beams also play the role of multiplexing optical information thus enhancing the security of photonic tally pair. When the photonic tally pair are brought into proximity with each other and illuminated by broadband white light, only vortex beams with specific topological charges and colors can form the pre-designed machine-recognizable patterns.

Figure S1(c) shows the design concept consisting of pairs of optical elements that form the basis of the photonic tally. The SPP is a well-known optical element to generate optical vortex carrying OAM, by introducing a spiral phase delay about its asymmetric central point. However, if the vortex beam passes through another SPP with the opposite topological charge and same orientation of lateral discontinuity, it will effectively resemble a beam passing through the combined glass slab with negligible fluctuation in amplitude and phase of incident beam (Fig. S1(c) inset). Analogous to the complementary inner morphology of the tiger tally pair, the SPP pair has complementary morphology, thus leading to counteracting OAM in momentum space and conjugated topological charges of phase. Vortex beam with spiral phase taking the form of $\exp(il\varphi)$ has the twisted wavefront as shown in Fig. S1(d), where $l$ represents topological charges of the vortex beam. The wavevector of a vortex beam is perpendicular to its equiphase surface, possessing the components along its propagating direction and azimuthal direction($l$), as indicated by the colorful arrows in Fig. S1(d) and Fig. S2. The azimuthal components of vortex wavevectors $k_\varphi$ are proportional to its OAM value and will counteract parts of each other when the optical vortices coalesce, indicated by the gradient red arrows. Thus, not only in real space does the complementary morphology of SPP pair extinguish the wavefront perturbation and abide by rectilinear propagation of light; but also in momentum space, as the angular momentum provides complementary features for the design of photonic tally pair.

**The coherence analysis of CVB units.**

The spatio-temporal coherence is represented by the complex degree of coherence $\gamma_{12}(\tau)$, which is defined by the mutual coherence function, $\Gamma_{12}(\tau) = \langle u_1(t+\tau) u_2^*(t) \rangle$, where the subscripts 1 and 2 stand for two source points to be investigated, $u$ is the corresponding wavefunction, e.g., electric field distribution on the observation screen(2). $\langle \cdots \rangle$ is the time average operator. $t$ is time, while $\tau$ is the time interval of electromagnetic wave arriving at the observation screen through these aforementioned two paths. Thus, the complex degree of coherence can be defined as $\gamma_{12}(\tau) = \Gamma_{12}(\tau) / \sqrt{\Gamma_{11}(0)\Gamma_{22}(0)}$, $\Gamma_{11}(0)$ and $\Gamma_{22}(0)$ are the light intensity at the observation screen coming from the point source 1 and 2. Usually, they are labelled by $I_1$ and $I_2$, respectively.

The degree of spatial coherence is defined as $\mu_{12} = \gamma_{12}(0) = \Gamma_{12}(0) / \sqrt{\Gamma_{11}(0)\Gamma_{22}(0)}$, $\Gamma_{12}(0)$ is the mutual light intensity. Usually, it's labelled by $J_{12}$. The degree of temporal coherence is defined as $\gamma(\tau) = \Gamma(\tau)/\Gamma(0)$. Since only one source point is considered when we investigate temporal coherence, the subscripts 1 and 2 are ignored. Furthermore, the complex degree of coherence can be decomposed into the product of the degree of spatial coherence and the degree of temporal coherence, as shown $\gamma_{12}(\tau) = \gamma(\tau)\mu_{12}$.

According to the van Cittert-Zernike theorem(2), the transverse spatial coherence width roughly equals to $\lambda_0 R/b$, where $\lambda_0$ is the operating wavelength, $R$ is the distance between the extended light source and the observation screen, and $b$ is the size of the extended light source. Taking the operating wavelength to be 500 nm, the distance $R$ to be 15 cm, and the size of a minimum aperture size of a common optical microscope to be 2.5 mm, the calculated transverse spatial coherence width roughly equals to 30 μm. Considering a shorter working wavelength and a larger aperture size of microscope lamp, we choose the diameter of our CVB unit to be 25 μm and make sure all CVB units work within the transverse spatial coherence width.

Consider the halogen lamp in the optical microscope as a black body, its coherence time is well defined as $\tau_c = \hbar/k_B T$, where $\hbar$ is the reduced Planck constant, $k_B$ is the Boltzmann constant, and $T$ is the temperature of the black body. Taking the temperature to be 3,000 K, the longitudinal temporal coherence length $L_c = c\tau_c$ should be around 1 μm, which is insufficient to generate colorful optical vortices. However, by adding the nanopillar-filter and narrowing the high transmission window in spectrum, the coherence length is enhanced to be several times. According to L. Mandel(*2*), the corresponding coherence length can be calculated as $L_c = \lambda_0^2/\Delta\lambda$, where $\lambda_0$ is the central wavelength, and $\Delta\lambda$ is the half width of transmission window. Taking the central wavelength to be 530 nm and the half width of transmission window to be 30 nm, the calculated longitudinal temporal coherence length is 9.36 μm. Considering a shorter working wavelength and a wider transmission window, we limit the topological charges of our CVB units within [-5, 5] and make sure the heights can meet the requirement of the longitudinal temporal coherence length.

In conclusion, our CVB units achieve sufficient spatio-temporal coherence and can be used directly under an incoherent white light source in a common optical microscope.

**Topological charge analysis of the CVB units under broadband white light illumination.**

We analyze the topological charges of the colorful vortex beam (CVB) units, elements of photonic tally, under both laser illumination and broadband white light. As shown in Fig. S7, we built a homemade setup to observe the optical micrograph of color vortex beam and its interference pattern with unconverted planewave. Under the illumination of halogen lamp, we can directly observe the doughnut shape intensity of colorful vortex beam as shown in Fig. S9 by using the pink light path in Fig. S7. With the modulus of topological charges creasing from 0 to 5, the radius of colorful rings in three color channels increases monotonously (Fig. S10). Even under illumination of broadband white light, the topological charges dependent colorful rings can be observed obviously by collimating the incident light, demonstrating the effectiveness and robustness of our CVB units. When changing the light source to lasers and detecting the interference signal using a CCD camera, the interference pattern between vortex beam (pink light path) and unaffected planewave (cyan light path) can be obtained, as shown in Fig. S11. By counting the number of interference fringes $n_{left}$ and $n_{right}$ in left and right regions, the difference value $n_{right} - n_{left}$ indicates the topological charge of incident vortex beam. With the topological charges changing from negative to positive, the residual fringes translocate from left to right. Both radii of colorful rings and fringes of colorful vortex beam in three different color channels manifest the topological charges are from -5 to 5, same as design.

To analyze the colorful optical vortices under broadband white light illumination, we examined intensity and phase profiles at the neighbor wavelengths region of 457 nm, 532 nm, 633 nm, topological charges within the transmitted windows can be further confirmed. As shown in Fig. S12, when the central wavelength of nanopillar color-filters match the design wavelengths, the intensity profiles remain doughnut shape in the whole range of high transmittance windows (Fig. S12(a)-(c)). Meanwhile, if we define the topological charges $q$ as phase wrapping along an arbitrarily closed loop $C$ in the doughnut shape intensity, $q = \frac{1}{2\pi}\oint_C \nabla\varphi(\mathbf{r}) \cdot d\mathbf{r}$, topological charges of the CVB units in transmitted windows remain to be 1 as same as the design wavelengths (Fig. S12(a)-(c)). However, when the central wavelength of nanopillar color-filters (633 nm) mismatches the design wavelength of the CVB unit surface (457 nm), either doughnut shape intensity or phase wrapping of 2π cannot maintain the *status quo*. Intensity profiles and phase wrapping of the CVB units with different topological charges are exhibited in Fig. S13 (a) to (d), the central wavelengths of these the CVB units are all 532 nm, while the topological charges are 1, 2, 3, and -1, respectively. Obviously, the ring diameters of doughnut shape intensity increase with topological charges and the phase wrapping along doughnut shape intensity loops validate the nominated topological charges. The sign of topological charges indicates the wrapping direction of phase as shown in Fig. S13 (a) to (d), agrees well with interference results in Fig. S11.

In summary, when examined on focal plane, both experimental and simulated results show the topological charges of the CVB units are conserved within corresponding high transmittance windows and the colored optical vortices are verified.

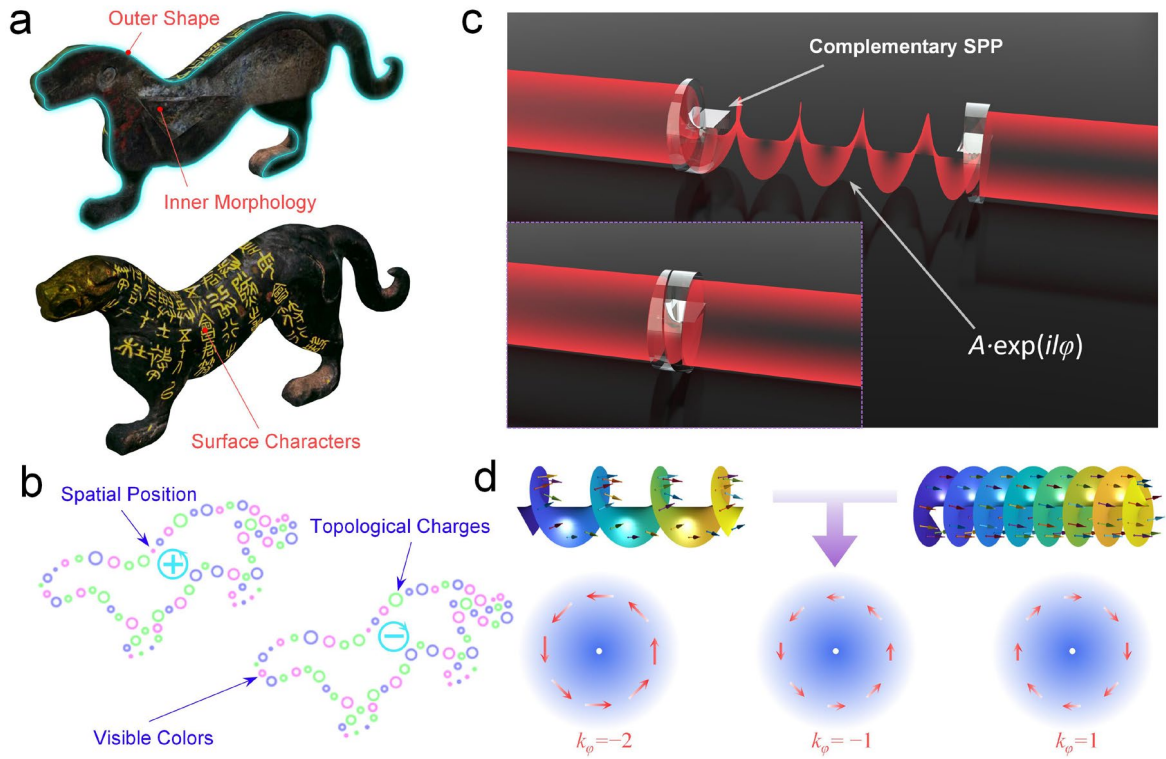

Figure S1. The analogy between "tiger tally" and "photonic tally". (a) Photograph of a tiger tally pair from ancient China, used as security tokens due to the three-dimensional complementary nature of outer shape, inner morphology, and surface characters. (b) Photonic tally pair, holding complementary features in topological charges, visible colors, and spatial positions. (c) Twisting and untwisting of wavefront after planewave passing through complementary SPP pair in real space. The topological charges of SPP pair are opposite and the orientations of lateral discontinuity are the same. Inset: merged SPP pair when they approach each other, the combined "plate" annihilates the wavefront twisting of single SPP. (d) Coalesce of optical vortices with different topological charges and cancellation of corresponding OAM.

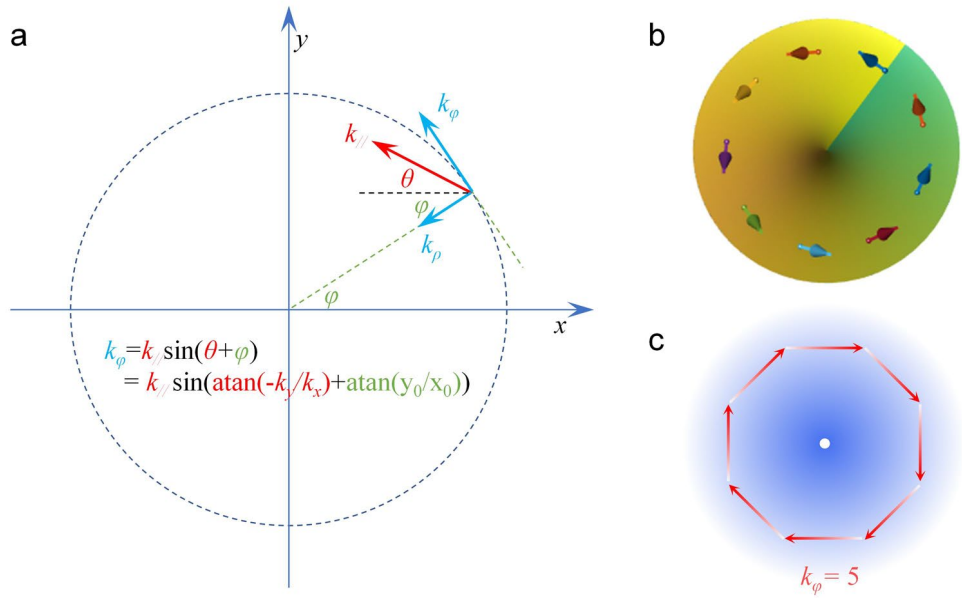

Figure S2. Azimuthal components of momentum in vortex beam. (a) Wavevector $k$ is firstly decomposed into vertical component $k_z$ along the propagation direction and parallel component $k_{//}$. And then, $k_{//}$ is decomposed into radial component $k_\rho$ and azimuthal component $k_\varphi$. (b) If the incident wave takes the form of $\exp(il\varphi)$, where $l$ represents the topological charges and $\varphi$ is the azimuthal angle, the radial component $k_\rho$ will disappear and topological charge $l$ equals to $k_\varphi$. Top view of the twisted wavefront shows there is no redundant radial momentum, the only lateral momentum is along azimuthal direction, this is also called angular momentum of vortex beam. (c) Azimuthal momentum of vortex beam, the gradient blue color stands for normalized value of azimuthal momentum, i.e., $|k_\varphi/k|$. When approaching the singularity of vortex beam, $|k_\varphi/k|$ approaches 1, and the propagation wave will lose its propagation component $k_z$ hence will be converted into evanescent wave at singularity. The gradient red arrows represent azimuthal momentum at a specific radius, the length of the arrows stands for the modulus of azimuthal momentum and the direction of the arrows stands for the direction of azimuthal momentum.

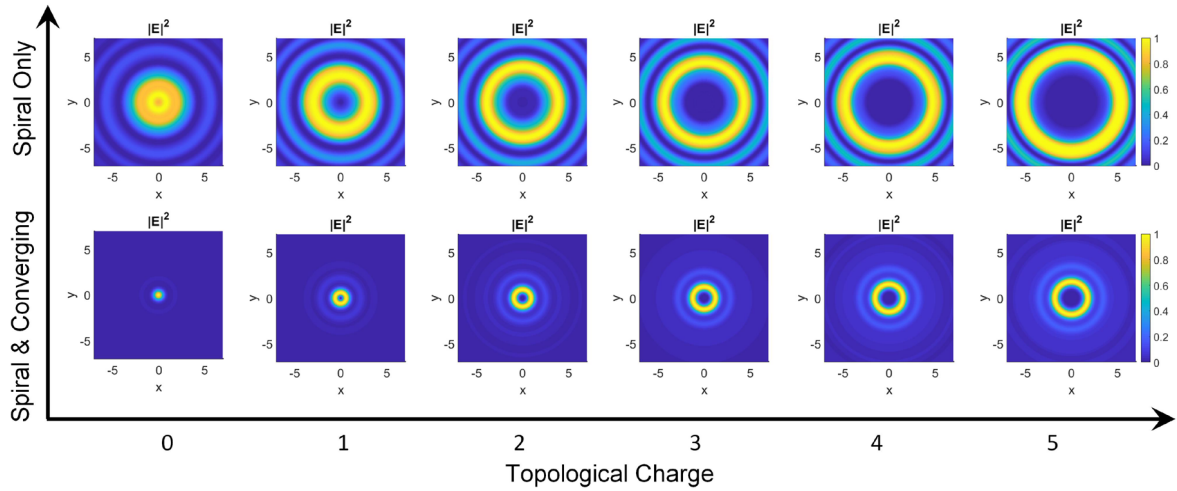

Figure S3. Comparison of purely spiral phase and the combination of spiral phase with converging phase. The light field at focal plane is calculated according to vectorial Rayleigh Sommerfeld diffraction integral. In all cases of topological charges increase from 0 to 5, the intensities with converging phases are more concentrated, leading to higher energy density and SNR.

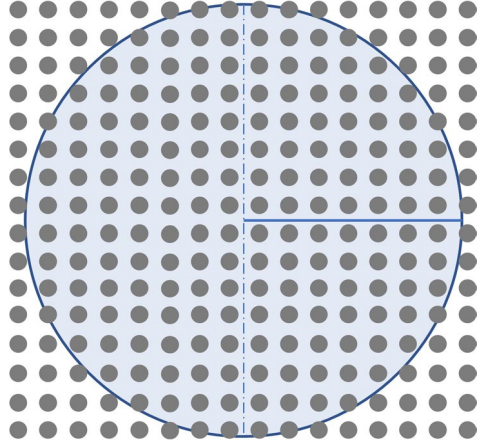

Figure S4. Nanopillar arrangement scheme for the CVB units. The circle contour stands for the radius of the CVB units, while blue full line stands for the lateral discontinuity of spiral phase plate. Firstly, we assume the nanopillar diameter is $d$ and the pitch is $p$, the radius of the CVB units is $R$. To avoid central singularity and lateral discontinuity, we arrange the nanopillars in grids and the coordinates can be given by the following equation,

$$x_n = \frac{2n \mp 1}{2} \cdot p, \quad -\left[\frac{R}{p}\right] \leq n \leq -1 \text{ or } 1 \leq n \leq \left[\frac{R}{p}\right]$$

$$y_m = \frac{2m \mp 1}{2} \cdot p, \quad -\left[\frac{R}{p}\right] \leq m \leq -1 \text{ or } 1 \leq m \leq \left[\frac{R}{p}\right]$$

where the sign $\mp$ depends on positive or negative values of $n$ and $m$. Meanwhile, the indices $n$ and $m$ should satisfy the following constrain,

$$\left[(2n \mp 1)p \pm d\right]^2 + \left[(2m \mp 1)p \pm d\right]^2 \leq 4R^2$$

to arrange nanopillars as more as possible and avoid diffraction effects caused by edges with overmuch symmetry.

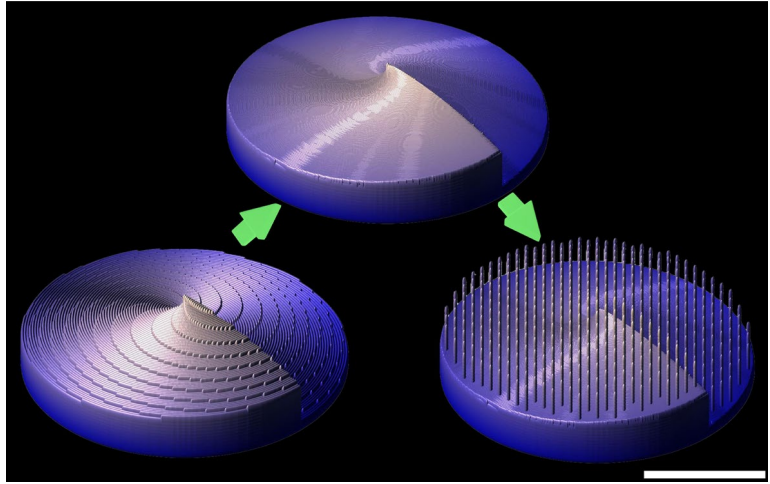

Figure S5. Generated 3D models for the two-density three-step nanoscale 3D printing, rough inner region, smooth top surface of the CVB units and adding on nanopillars, the scale bar is 10 μm.

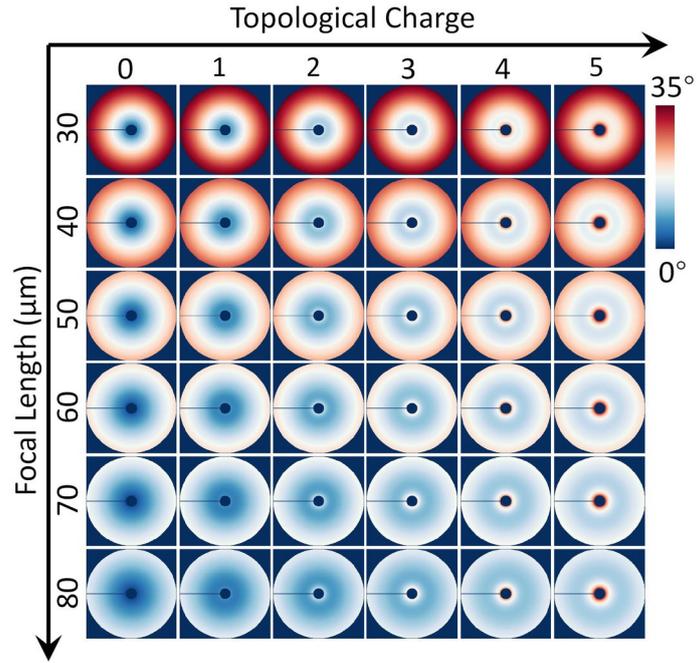

Figure S6. Detailed figure of Figure 2(c) inset. Local surface angles of the CVB units with respect to different focal lengths and topological charges. The diameters of the CVB units are fixed at 25 μm.

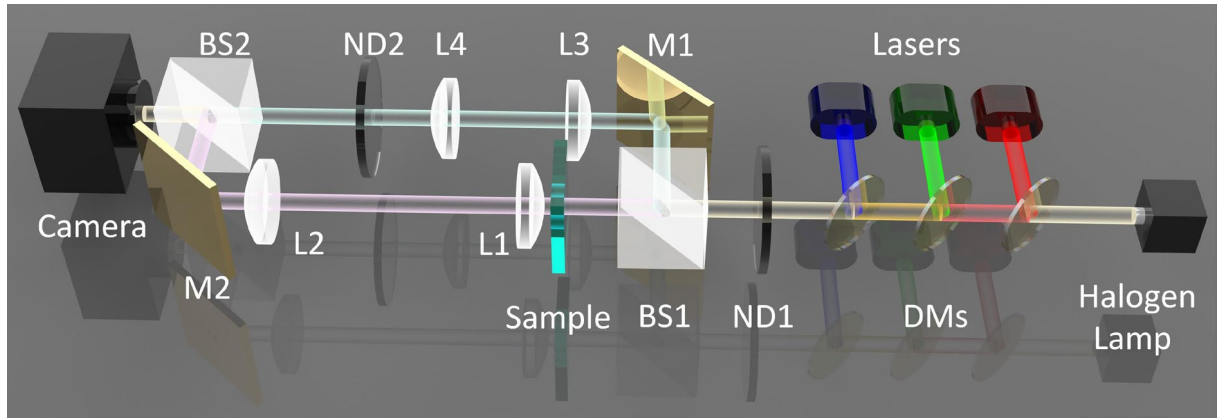

Figure S7. Homemade optical setup for colorful optical vortices measurement. Optical micrographs of the CVB units and doughnut shape intensity profiles can be measured via the pink optical path. While the corresponding topological charges can be examined through the spatial interference fringes between the pink and cyan optical paths. BS: beam splitter, M: mirror, ND: neutral density filter, L: lens.

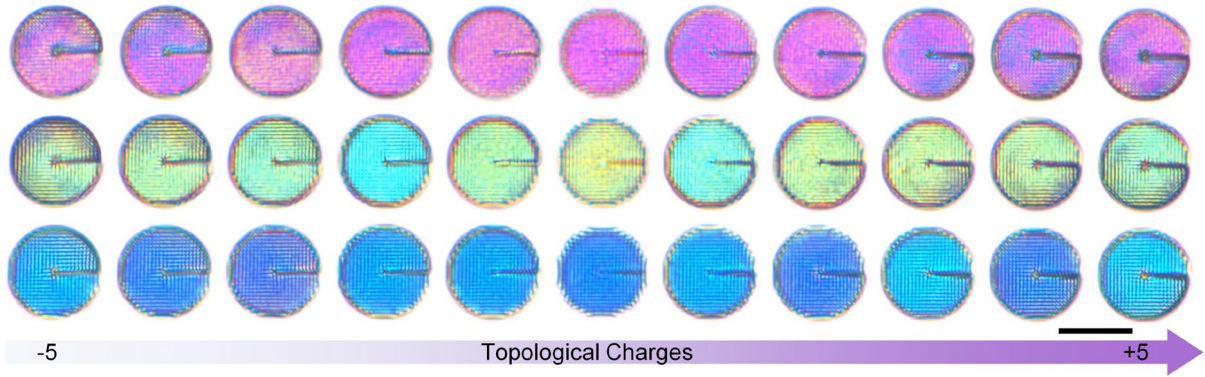

Figure S8. Optical micrographs of the CVB units in three color channels with topological charges ranging from -5 to 5. The scale bar is 20 μm.

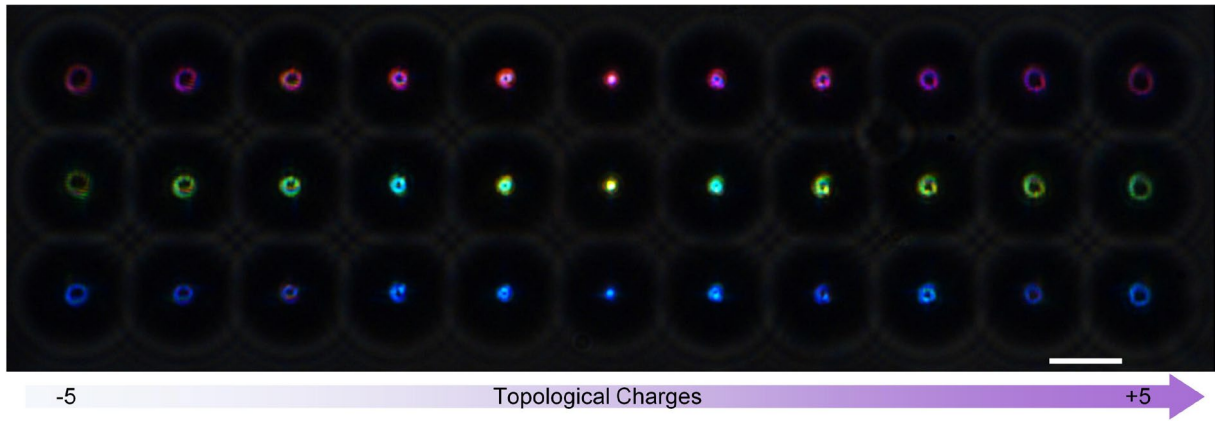

Figure S9. Doughnut shape intensity profiles of the CVB units in three color channels with topological charges ranging from -5 to 5. The scale bar is 20 μm.

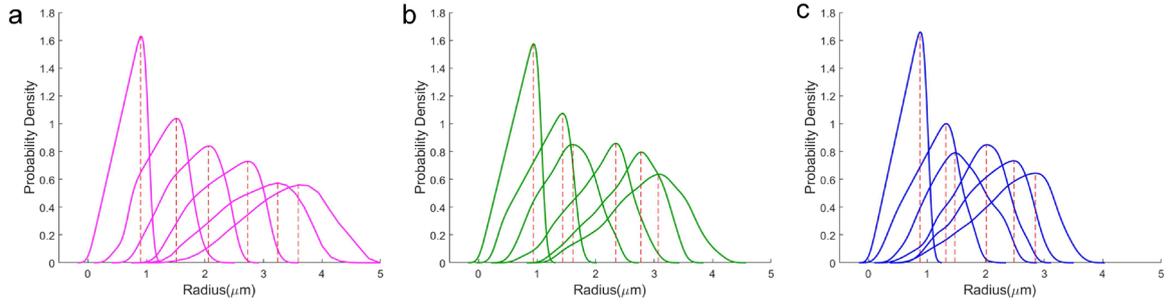

Figure S10. Probability density of colorful vortex rings for colorful the CVB units with topological charges from 0 to 5. The probability density curves are calculated from optical micrograph of focused colorful rings under a 50× objective lens with NA=0.4. Color pixels lighted up on CCD camera are recorded regarding to their radius and distance from the averaged center, and then are converted to probability density curves. Each curve stands for a CVB unit with specific topological charge and nanopillar height. With the increasing of topological charges, their peak positions of probability density curves move outer. This demonstrates that the radius of colorful rings increases with topological charges in all three RGB channels.

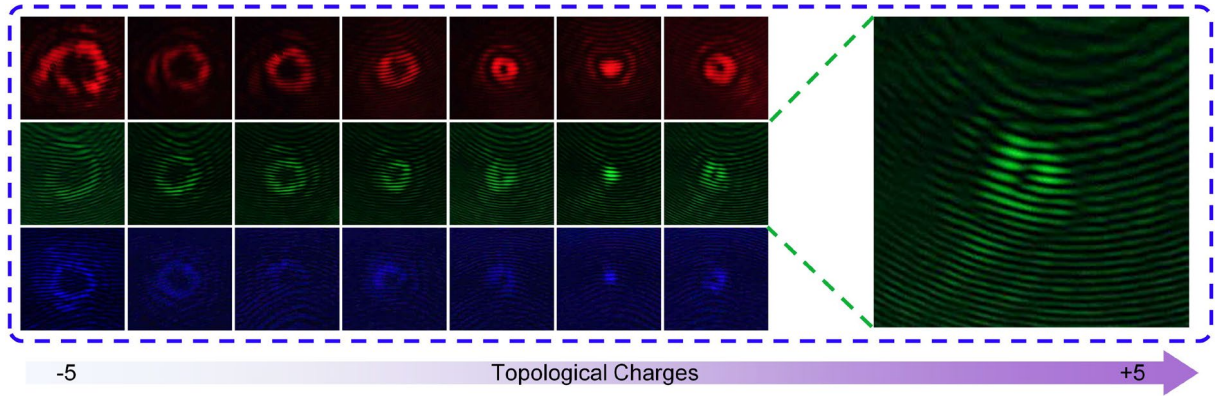

Figure S11. Interference patterns of the CVB units with topological charges from -5 to 1 in three color channels. Inset: enlarged interference pattern of green vortex beam with topological charges of 1.

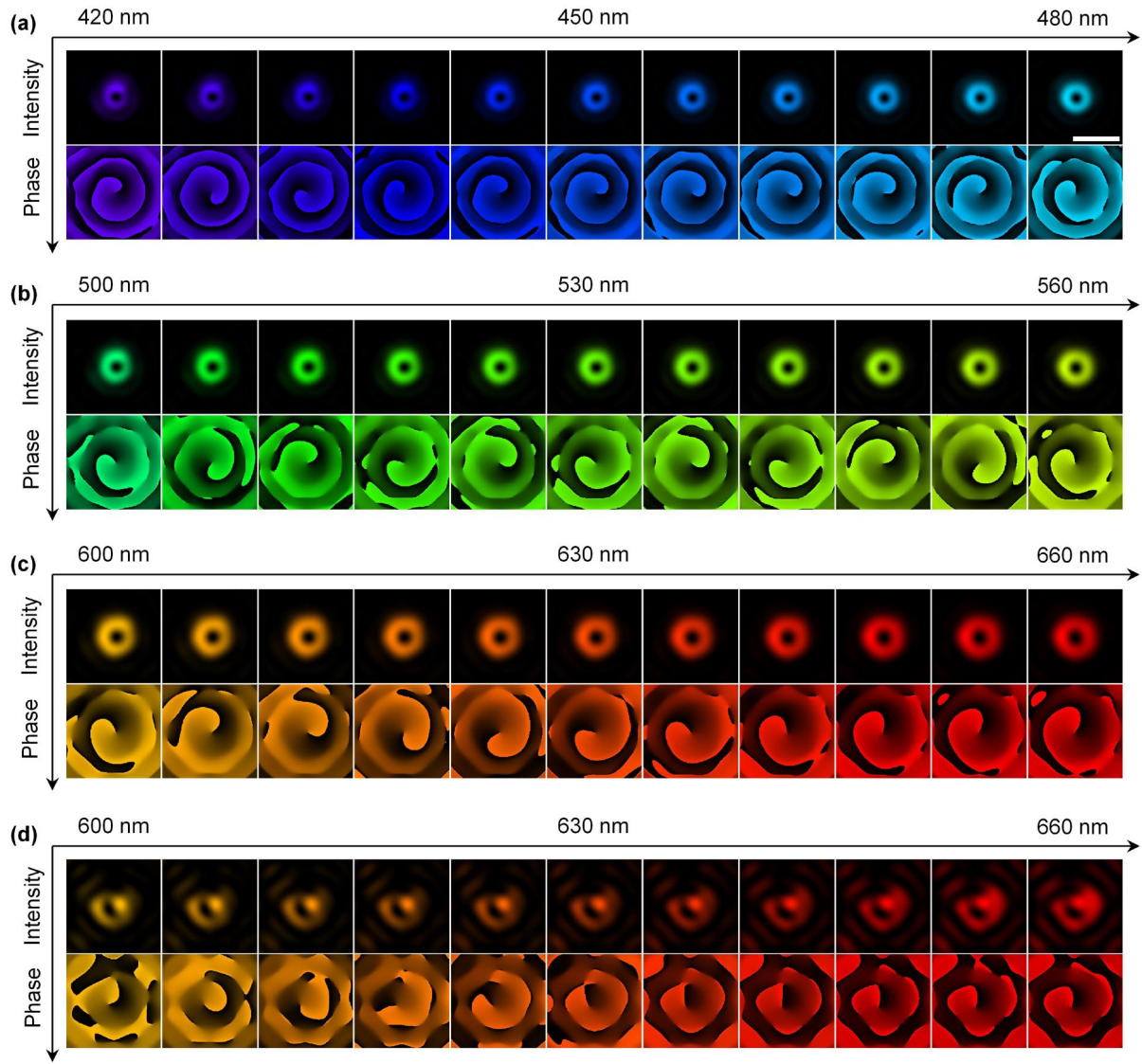

Figure S12. (a)-(c) Simulated intensity profiles and phase wrapping of the CVB units when the design wavelengths of the CVB units surface tomography match the high transmittance windows of nanopillars. (d) Simulated intensity profiles and phase wrapping of the CVB units when the design wavelength of the CVB units surface tomography is 457 nm while the color of corresponding nanopillars is red. The scale bar is 5 μm.

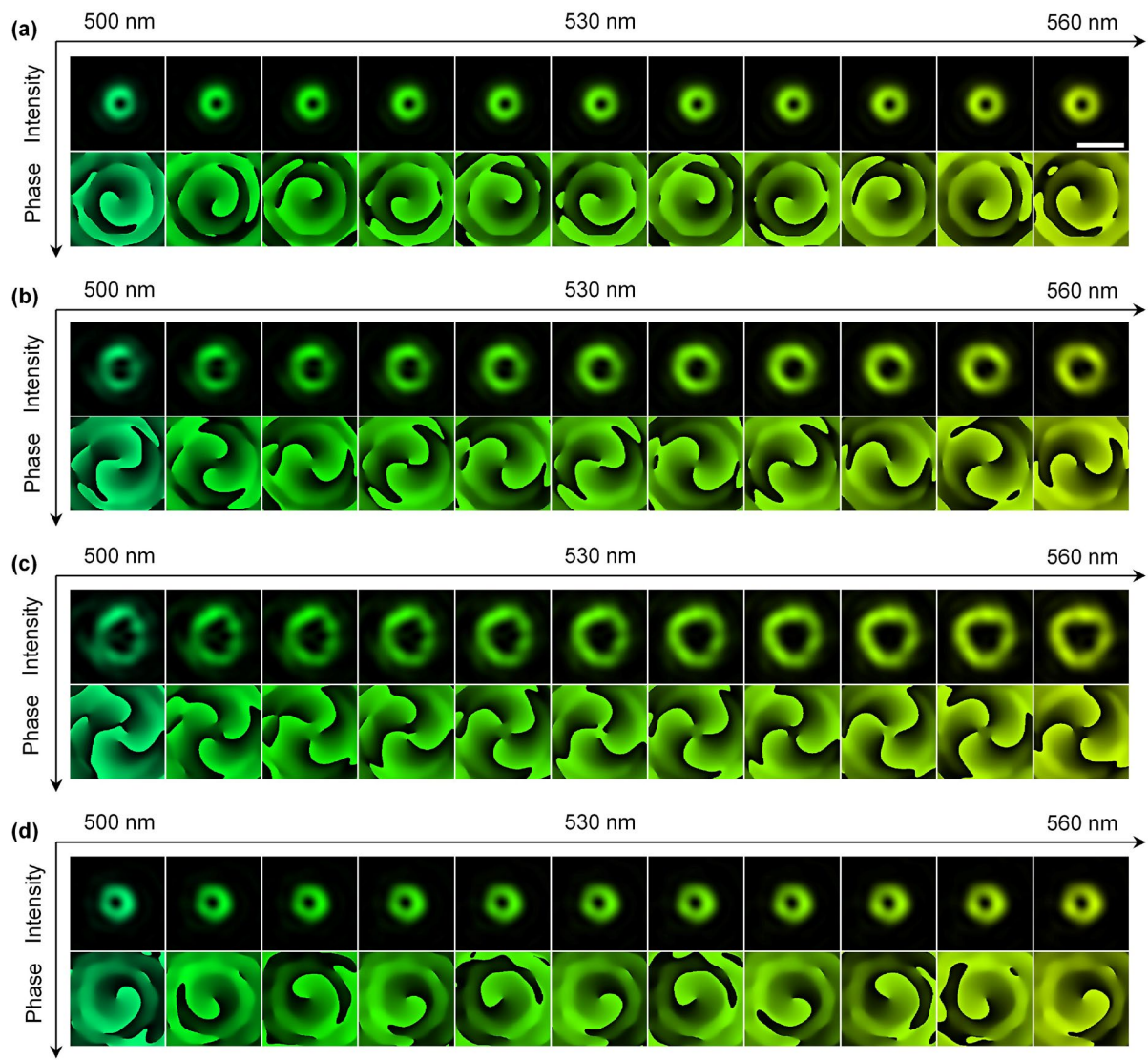

Figure S13. Simulated intensity profiles and phase wrapping of the green CVB units with topological charges 1, 2, 3, and -1. The scale bar is 5 μm.

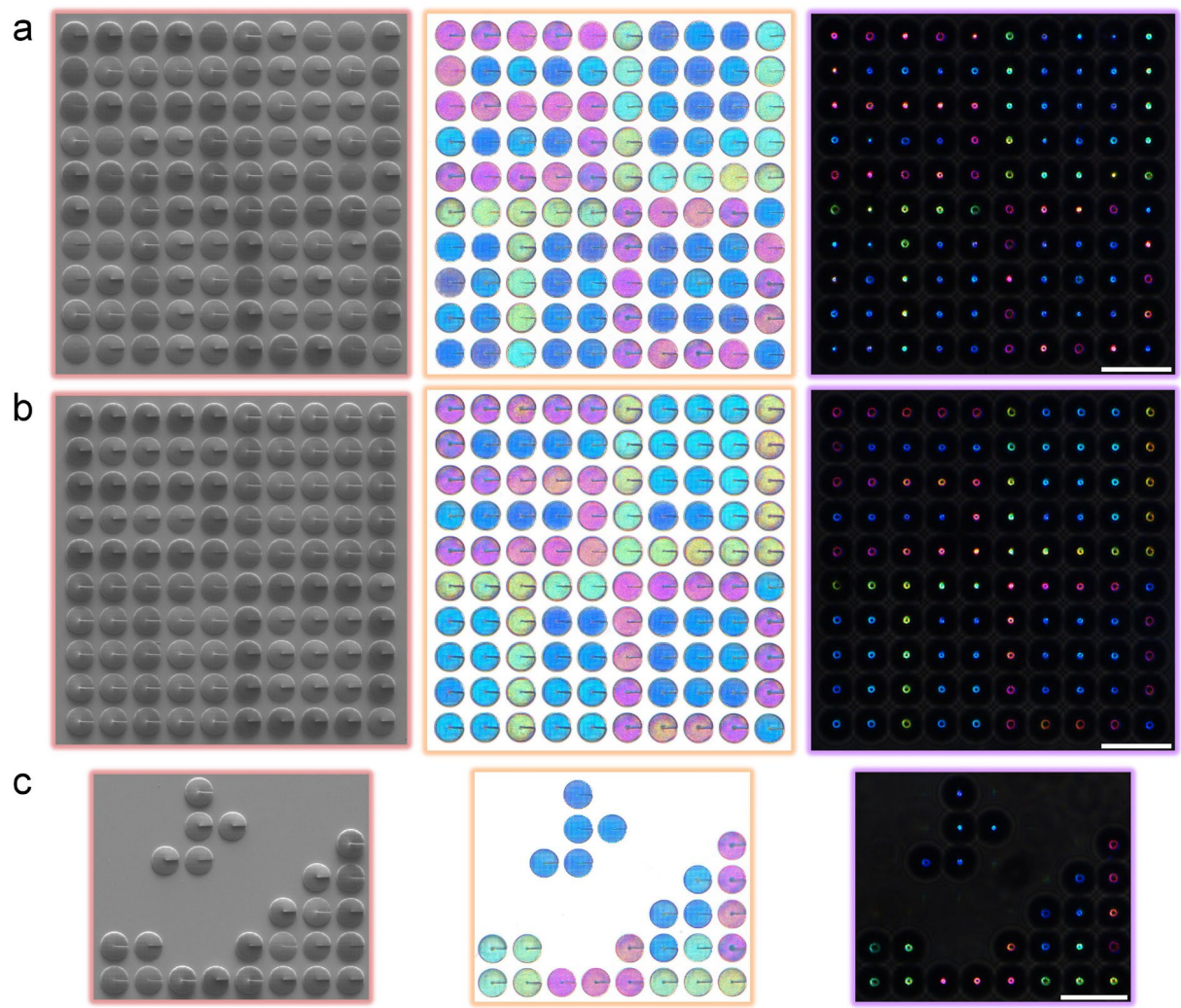

Figure S14. Origin images for Figure 3 in the main text. All scale bars are 50 μm.

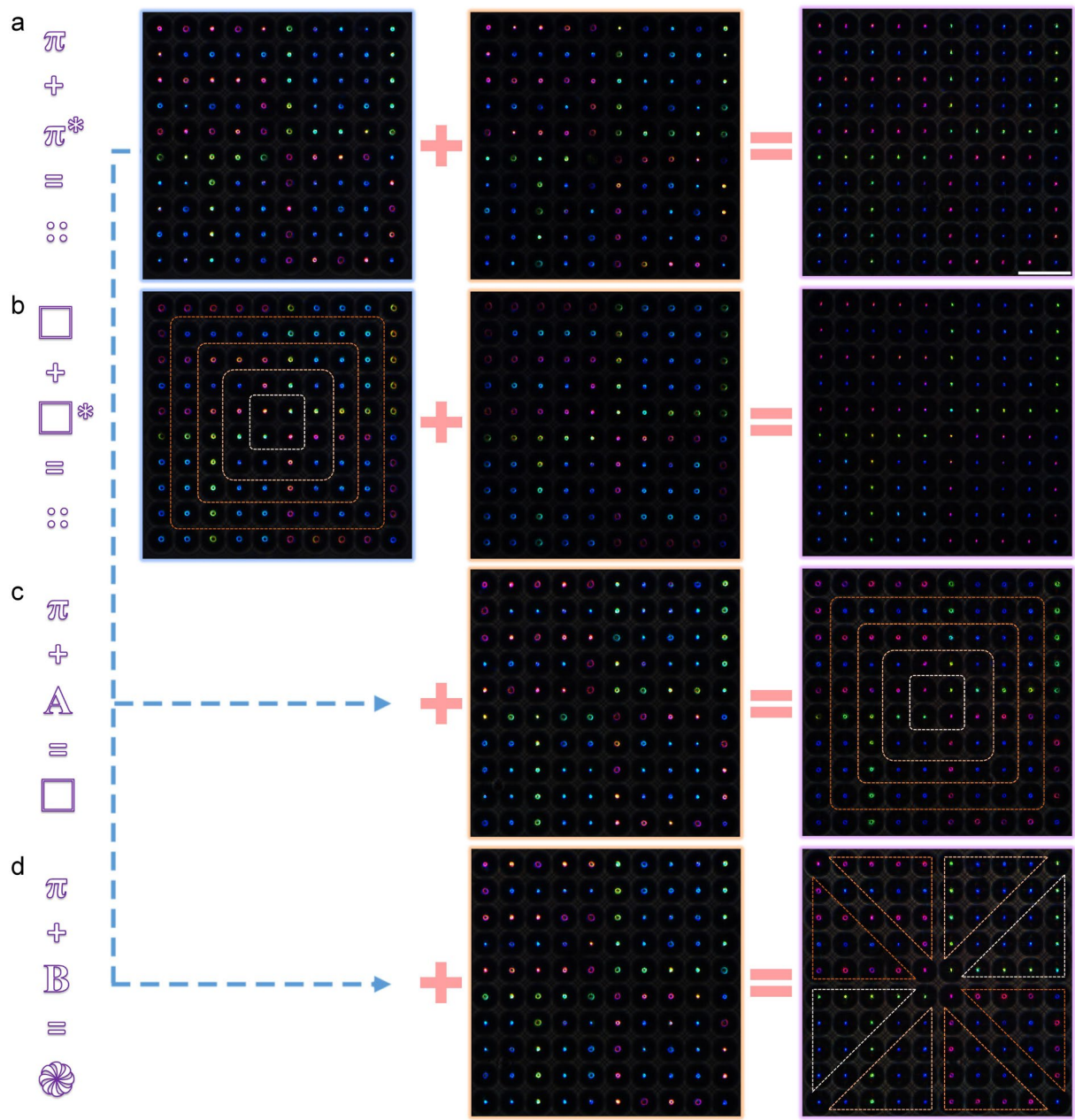

Figure S15. Origin images for Figure 4 in the main text. The scale bar is 50 μm.

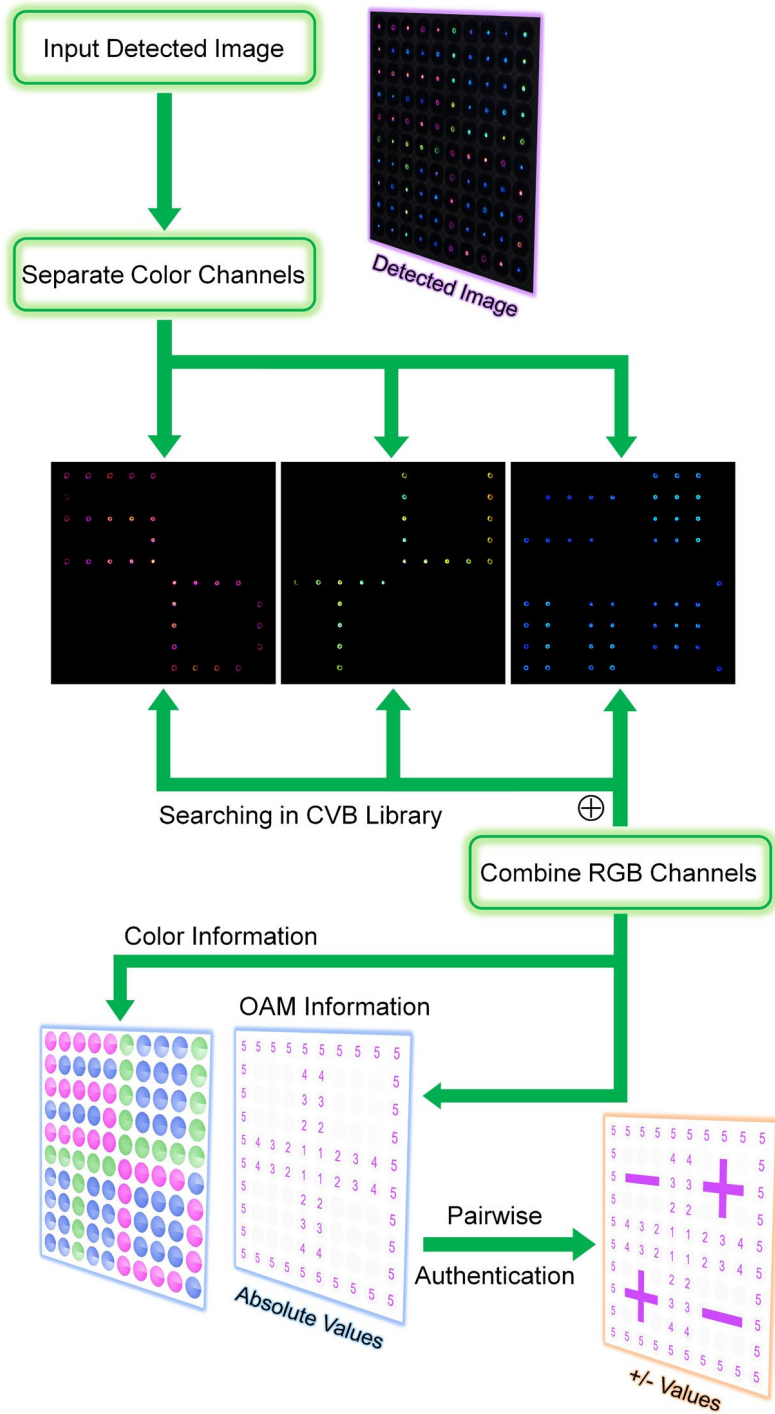

Figure S16. The flowchart of CVB detection and pairwise authentication.